\magnification=\magstephalf
\hyphenation{brem-sstrahlung}
\hsize=6.5truein 

\vsize=9.15truein
\hoffset=-0.1truein
\voffset=-.15truein 

\null

\baselineskip=24truept plus 0truept minus 0truept

\parindent=3em 


\overfullrule=0pt
\headline={\ifodd\pageno\rightheadline\else\rightheadline\fi}
\def\numbers{\def\rightheadline{\hfil\tenrm\folio\hfil}} 

\def\nonumbers{\def\rightheadline{\hfil}} 

\nonumbers 

\nopagenumbers 

\tolerance=800 \hbadness=5000
\newdimen\listindent\listindent=2em 

\null

\def\gapprox{\lower.4ex\hbox{$\;\buildrel >\over{\scriptstyle\sim}\;$}}
\def\lapprox{\lower.4ex\hbox{$\;\buildrel <\over{\scriptstyle\sim}\;$}}

\def\p{\raise.45ex\hbox{$\wp$}}
\def\Mdot{\dot M}

\def\ME{\dot M_{_{\rm E}}}
\def\LE{L_{_{\rm E}}}
\def\msun{M_\odot}
\def\mrate{\,\msun{\rm yr}^{-1}}

\def\lambdacoh{\lambda_{\rm coh}}
\def\lambdaii{\lambda_{\rm ii}}
\def\lambdacyc{\lambda_{\rm L}}
\def\etaff{\eta_{\rm ff}}
\def\etahyb{\eta_{\rm hyb}}

\def\sqr#1#2{{\vcenter{\vbox{\hrule height.#2pt
     \hbox{\vrule width.#2pt height#1pt \kern#1pt
         \vrule width.#2pt}
    \hrule height.#2pt}}}}

\def\refs{\leftskip=2em\parindent=-2em}


\vskip1.5truein
\centerline{\bf ION VISCOSITY MEDIATED BY TANGLED MAGNETIC FIELDS:}
\centerline{\bf AN APPLICATION TO BLACK HOLE ACCRETION DISKS}
\bigskip
\bigskip
\centerline{Prasad Subramanian$^1$, Peter A. Becker,$^2$
and Menas Kafatos$^2$}
\bigskip
\centerline{Center for Earth Observing and Space Research,}
\centerline{Institute for Computational Sciences and Informatics,}
\centerline{George Mason University, Fairfax, VA 22030-4444}
\centerline{$^1$psubrama@gmu.edu}
\centerline{$^2$also Department of Physics and Astronomy,}
\centerline{George Mason University, Fairfax, VA 22030-4444}
\bigskip
\bigskip
\bigskip
\bigskip
\vskip2.0truein
\centerline{(submitted to the {\it Astrophysical Journal})}
\vskip 3.0truecm
\vfil
\eject

\def\t0{{\theta_0}}

\def\r0{R_0}

\def\sig{\sigma_{_{\rm T}}}

\numbers
\pageno=2
\bigskip
\centerline{ABSTRACT}

We examine the viscosity associated with the shear stress exerted
by ions in the presence of a tangled magnetic field. As
an application, we consider the effect of this mechanism on
the structure of black hole accretion disks. We do not attempt
to include a self-consistent description of the magnetic field.
Instead, we assume the existence of a tangled field with coherence
length $\lambdacoh$, which is the average distance between the
magnetic ``kinks'' that scatter the particles. For simplicity,
we assume that the field is self-similar, and take $\lambdacoh$
to be a fixed fraction of the local radius $R$. Ion viscosity in the presence of magnetic fields is generally taken to be the cross-field viscosity, wherein the effective mean free path is the ion Larmor radius $\lambdacyc$, which is much less than the ion-ion Coulomb mean free path $\lambdaii$ in hot accretion disks. However, we arrive at a
formulation for a ``hybrid'' viscosity in which the tangled magnetic
field acts as an intermediary in the transfer of momentum between
different layers in the shear flow. The hybrid viscosity greatly
exceeds the standard cross-field viscosity when
$(\lambda / \lambdacyc) \gg (\lambdacyc / \lambdaii)$, and
$\lambda = (\lambdaii^{-1} + \lambdacoh^{-1})^{-1}$ is the
effective mean free path for the ions.
This inequality is well satisfied in hot accretion disks, which
suggests that the ions may play a much larger role in the
momentum transfer process in the presence of magnetic fields
than was previously thought. The effect of
the hybrid viscosity
on the structure of a steady-state, two-temperature, quasi-Keplerian
accretion disk is analyzed, and the associated Shakura-Sunyaev $\alpha$
parameter is found to lie in the range $0.01 \lapprox \alpha \lapprox 0.5$.
The hybrid viscosity is influenced by the degree to which the magnetic
field is tangled (represented by the parameter $\xi \equiv \lambdacoh/R$),
and also by the relative accretion rate $\Mdot/\ME$, where $\ME\equiv \LE/c^2$
and $\LE$ is the Eddington luminosity. When the accretion rate is
supercritical ($\Mdot/\ME \gapprox 1$), the half-thickness of
the disk exceeds the local radius in the hot
inner region and vertical motion becomes important. In such cases the
quasi-Keplerian model breaks down, and the radiation
viscosity becomes comparable to the hybrid viscosity.

\vfil
\eject

\bigskip
\centerline{SUBJECT HEADINGS}
Accretion disks, Ion viscosity, Plasma viscosity, MHD turbulence,
 Magnetic fields: tangled, Black hole physics.

\bigskip
\centerline{\bf 1. INTRODUCTION}
\bigskip

\bigskip
\centerline{1.1. \it Background}
\bigskip

Viscosity in accretion disks around compact objects has been the
subject of investigation for nearly 20 years (for a review, see
Pringle 1981). It was recognized very early on that ordinary molecular
viscosity cannot produce the level of angular momentum transport
required to provide accretion rates commensurate with the observed
levels of emission in active galaxies, quasars, and galactic black-hole
candidates (Shakura \& Sunyaev 1973). Consequently, the actual nature of
 the microphysics
leading to viscosity in such flows has been the subject of a great deal of
speculation. For plane-parallel flows with shear velocity
$\vec u = u(y)\,\hat z$, the shear stress is defined as the flux of
$\hat z$-momentum in the $\hat y$-direction. In lieu of a detailed
physical model for the process, the work of Shakura \& Sunyaev (1973)
led to the embodiment of all the unknown microphysics into a single
parameter $\alpha$, defined by writing the shear stress as
$$
\alpha P \equiv - \eta {du \over dy} = {3 \over 2}\,\eta\,\Omega_{\rm kepl} \,,
\eqno(1.1)
$$
where $P$ is the total pressure, $\eta$ is the dynamic viscosity, and
 $\Omega_{\rm kepl}$
is the local orbital frequency inside a quasi-Keplerian accretion disk.
Note the appearance of the negative sign,
which is required so that $\eta$ is positive-definite. Order-of-magnitude
arguments advanced by
Shakura \& Sunyaev (1973) lead to the general conclusion that $0 < \alpha
< 1$. This stimulated the development of a large number of theoretical models
in which $\alpha$ is treated as a free parameter; in many of these models
$\alpha$ is taken to be a constant. This has been partially motivated by
the fact that in quasi-Keplerian accretion disks around black holes,
observational quantities like the luminosity depend only weakly upon
$\alpha$. This enabled  progress to be made without precise knowledge
of the microphysical viscosity mechanisms. However, this does not eliminate
the need for an understanding of these mechanisms, and without such an
understanding, much of the high temporal resolution data being collected
by space instrumentation cannot be fully interpreted. Several processes
have been suggested to explain the underlying microphysical viscosity mechanism.
Initial developments focused on the turbulent viscosity first proposed
by Shakura \& Sunyaev (1973), and later investigated more rigorously by
Goldman \& Wandel (1995). Although the presence of turbulence in accretion
disks is probably inevitable, it is unclear whether this particular
viscosity mechanism will dominate over other processes that may be
operating in the same disk, such as radiation viscosity (Loeb \& Laor 
1992), magnetic viscosity (Eardley \& Lightman 1975), and ion viscosity
(Paczynski 1978, Kafatos 1988).

The paper is organized as follows. In \S 1.2 we provide a general introduction
 to ion viscosity in accretion disks. In \S 1.3 we give a heuristic derivation
 of ion viscosity in the absence of magnetic fields. In \S 1.4 we discuss
 cross-field ion viscosity in the presence of magnetic fields. In \S 2 we 
derive  the hybrid viscosity due to ions in the presence of tangled
 magnetic fields for the general case of a plane-parallel shear flow. We 
apply our results to two-temperature accretion disks in \S 3. The disk
 structure equations are outlined in \S 3.1 and in \S 3.2 we discuss the 
main results. we discuss the main conclusions in \S 4.
 
\bigskip
\centerline{1.2. \it Ion Viscosity}
\bigskip

Ion (plasma) viscosity in accretion flows has been previously investigated
by Paczynski (1978), Kafatos (1988), and Filho (1995). In this process,
angular momentum is transferred between different layers in the shear
flow by ions that interact with each other via Coulomb collisions. The
mean free path for the process is then the Coulomb mean free path. Few
detailed astrophysical models have been constructed using the plasma viscosity
as the primary means for angular momentum transport because of the presumed
sensitivity of this mechanism to the presence of magnetic fields. The effect
of the magnetic field is particularly important when the ion gyroradius is
less than the Coulomb mean free path {\it and} the orientation of the local
field is perpendicular to the local velocity gradient, because in this case
different layers in the shear flow cannot communicate effectively. This point
was first raised by Paczynski (1978), who argued that even for very weak fields
(as low as $10^{-7}\,$G), this effect is enough to almost completely quench
the ion viscosity. This is problematic, since it is very reasonable to expect
near-equipartition magnetic fields to be present in an accretion flow, with
strengths many orders of magnitude greater than $10^{-7}\,$G.

Implicit in Paczynski's argument is the assumption that the local magnetic
field is {\it exactly perpendicular} to the local velocity gradient. However,
near-equipartition magnetic fields would probably be tangled over macroscopic
length scales, as evidenced, for example, by simulations of the nonlinear stage
of the Balbus-Hawley instability (Matsumoto et al. 1995). If this were not so
and the magnetic field was ordered over macroscopic length scales, it would
 imply
that the flow dynamics are dominated by the magnetic field, which contradicts
 the
assumption of equipartition between the magnetic and kinetic energy densities.
We argue below that the presence of {\it tangled} magnetic fields effectively
eliminates Paczynski's concern, because ions are able to transfer a significant
fraction of their momentum by travelling along field lines connecting two 
different
layers in the shearing plasma.

\bigskip
\centerline{1.3. \it Field-Free Coulomb Viscosity}
\bigskip

Consider a field-free plasma with Coulomb mean free path $\lambdaii$ and
shear velocity distribution $\vec u = u(y)\,\hat z$, where we set $u(0)=0$
without loss of generality. The shear stress
is equal to the net flux of $\hat z$-momentum in the $\hat y$-direction.
In terms of the field-free dynamic viscosity $\etaff$ the shear stress
is given by
$$
- \etaff\,{du\over dy} \equiv - N_i\,\sqrt{k T_i\over 2 \pi m_i}
\cdot m_i\,{du\over dy}\,\lambdaii \cdot 2\,, \eqno(1.2)
$$
where $N_i$ is the ion number density, $m_i$ is the ion mass, and
$T_i$ is the ion temperature (Mihalas \& Mihalas 1984). The first factor on
the right-hand side of equation~(1.2) represents the unidirectional
particle flux crossing the $y=0$ plane, and the second factor
is the magnitude of the average $\hat z$-momentum carried by particles
originating a mean distance $\lambdaii$ from the plane. The factor of
$2$ accounts for the transport of particles in both directions across
the plane. For pure, fully-ionized hydrogen, we have
$$
\lambdaii = v_{\rm rms} t_{ii} = 1.8 \times 10^5{T_i^2 \over N_i
\ln\Lambda}\,, \eqno(1.3)
$$
where $\ln \Lambda$ is the Coulomb logarithm, $v_{\rm rms} = \sqrt{3kT_i/ m_i}$
is the root mean square velocity of the Maxwellian distribution, and
$$
t_{ii} = 11.4\,{T_i^{3/2} \over N_i \ln \Lambda} \eqno(1.4)  
$$
is the mean time between Coulomb collisions. This yields the standard result
for the field-free dynamic viscosity obtained by Spitzer (1962),
$$
\etaff = 2.2 \times 10^{-15}\,{T_i^{5/2} \over \ln \Lambda}\ \ 
{\rm g \, cm^{-1} \, s^{-1}}\,. \eqno(1.5)
$$

Equation~(1.5) is valid provided the gas is collisional, which in
this case requires that the mean free path of the protons $\lambdaii$
be much smaller than any macroscopic length scale in the problem. It
turns out, however, that for gas accreting onto a black hole,
$\lambdaii/R$ can exceed unity in general, where $R$ is the local
radius. In this case, regions that are separated by distances larger
than the characteristic length over which the velocity varies
[$v/(d v/dR)\sim R$] can easily exchange particles and therefore
momentum as well. In such ``non-local'' situations, the shear stress
is no longer simply proportional to the local velocity gradient, and
one must solve the full Boltzmann equation in order to study the
dynamics of the flow. Another problem that arises when
$\lambdaii/R \gapprox 1$ involves the shape of the ion velocity
distribution. When the ions are not effectively confined to
a small region of the flow with characteristic dimension $L \ll R$,
the local velocity distribution can become distinctly non-Maxwellian
due to the influence of processes occurring far away in the disk.
In such circumstances, the very existence of the ion temperature must
be called into question.

If this were the whole story, then the construction of disk models
using ion viscosity would present formidable challenges. However,
so far we have completely neglected the effects of the near-equipartition,
tangled magnetic field likely to be present in an actual accretion disk.
As we argue below, the presence of such a field will completely alter the
conclusions reached above if the coherence length of the field is much less
than the local radius $R$, because then the ions will be effectively confined
to a region of plasma with characteristic size $L \ll R$.

\bigskip
\centerline{1.4. \it Cross-Field Coulomb Viscosity}
\bigskip

Next we consider the shear stress exerted by ions inside a plasma containing
a magnetic field oriented in the $\hat z$-direction and moving with velocity
$\vec u = u(y)\,\hat z$, where $u(0)=0$. Hence the magnetic field is
exactly perpendicular to the local velocity gradient. In hot accretion disks, 
one
generally finds that $\lambdacyc \ll \lambdaii$ for near-equipartition magnetic
fields (Paczynski 1978), where
$$
\lambdacyc = 0.95 \, T_{i}^{1/2} B^{-1} \,, \eqno(1.6)
$$
is the Larmor radius of the ions in the presence of a magnetic field B. The 
shear stress is therefore given by
$$
- \eta_\perp\,{du\over dy} \equiv - 2 \cdot N_i\,\sqrt{k T_i\over 2 \pi m_i}
\cdot m_i\,{du\over dy}\,\lambdacyc \cdot
{\lambdacyc\over\lambdaii}\,, \eqno(1.7)
$$
where $\eta_\perp$ is the cross-field viscosity. This is similar
to equation~(1.2), except that the magnitude of the
average $\hat z$-momentum carried by particles crossing the plane
is now $\sim (du / dy)\lambdacyc m_i$ because the particles originate at a
mean distance $\sim \lambdacyc$ from the plane. Another modification is the
addition of the factor $(\lambdacyc/\lambdaii)$ which accounts approximately for the efficiency of the momentum transfer process. To
understand the efficiency factor, imagine an ion originating on the right
side of the plane, and spiraling about a magnetic field line. During one
gyration, the particle crosses from the right side of the plane to the left
side. Since $\lambdacyc \ll \lambdaii$ by assumption, the probability that
the particle will experience a Coulomb collision with another ion before
returning to the right side is $\sim \lambdacyc/\lambdaii$. Hence this factor
gives the mean efficiency of the momentum transfer process. The
cross-field viscosity can also be written as
$$
\eta_\perp = \etaff \left(\lambdacyc\over\lambdaii\right)^2
= 6.11 \times 10^{-26} \, {N_{i}^{2} \ln \Lambda \over T_{i}^{1/2} B^{2}} \,, \eqno(1.8)
$$
This expression agrees with the result for this case given by Kaufman (1960),
 to within a factor of the order of unity. We attribute the discrepancy to the 
approximate nature of our efficiency factor $(\lambdacyc/\lambdaii)$, which 
does not take several details like the pitch angle of the spiralling ions into
 account, and to the fact that we take the ions to be originating exactly at
 a distance $\lambdacyc$ away.

Since $(\lambdacyc/\lambdaii)^2 \ll 1$ even for
field strengths as low as $10^{-7}\,$G, Paczynski (1978) concluded
that the ion viscosity
plays a negligible role in determining the disk structure unless the
magnetic field essentially vanishes. However, Paczynski's conclusion
relies upon the assumption that the magnetic field is exactly
perpendicular to the local velocity gradient. We do not believe
that this assumption is justified when the magnetic field is
created dynamically within the disk, rather than imposed from the
outside. When the field is created dynamically in turbulent plasma,
numerical simulations indicate that the direction of the field varies
randomly in time and space (Matsumoto et al. 1995). In such situations,
the field is {\it tangled}, and it is more useful to consider a new,
``hybrid'' viscosity, where the effective mean free path is limited by
the coherence length of the magnetic field. We present a derivation
of the hybrid viscosity in \S~2, culminating with the expression for
 $\eta_{\rm hyb}$ in equation (2.14).

\bigskip
\centerline{\bf 2. ION VISCOSITY IN THE PRESENCE OF A TANGLED
MAGNETIC FIELD}
\bigskip

In \S~1.4, we considered the case of a shearing plasma
containing a magnetic field oriented in the $\hat z$-direction, exactly
perpendicular to the local velocity gradient. In an actual accretion
disk, we do not expect this to be the case very often. Instead, the
direction of the field is likely to be a random function of position
on scales exceeding the correlation length of the tangled magnetic field, which arises from MHD turbulence. It is therefore interesting to consider the shear
stress exerted by ions in the general case of a randomly directed field.
If $\lambdacyc \ll \lambdaii$, then we expect that ions moving between
different layers in the fluid will spiral tightly around the field
lines, in which case two of the components of the ion momentum are
obviously not conserved. On the other hand, the component of the ion
momentum {\it parallel} to the magnetic field {\it is} conserved
until the particle either experiences a Coulomb collision with another
ion or encounters an irregularity in the magnetic field. Hence the
transfer of momentum from one layer to another occurs via the component
of the particle momentum parallel to the magnetic field, and in this
sense the particles act like beads sliding along a string, in what is 
commonly referred to as the ideal MHD approximation.

The irregularities that scatter the ions may appear as either
stationary ``kinks'' or fast, short-wavelength electromagnetic waves
depending on the details of the turbulence. If the
particles interact with the field primarily via wave-particle
scattering, then the waves must be explicitly included as
a dynamical entity in the momentum transfer process. In fact,
the shear stress due to the waves themselves may dominate the
situation if the wave energy density surpasses that of the
particles. However, such large wave energy densities cannot
be created if the field is generated dynamically within the
plasma, as we assume here. Furthermore, the relatively fast-moving
ions that carry momentum in our picture will not often encounter
short-wavelength electromagnetic waves with sufficient amplitude
to scatter them very strongly. Conversely, the ions {\it will} be
strongly scattered by encounters with long-wavelength, slow-moving
kinks in the magnetic field. We therefore ignore the dynamical
consequences of the fast waves, and treat the irregularities as
stationary kinks. We will elaborate on this aspect in \S 4.

If the field is frozen into the plasma, then the ion momentum
will ultimately be transferred to the local gas via either Coulomb
collisions or encounters with magnetic irregularities. The probability
per unit length for either type of interaction to occur is proportional
to the reciprocal of the associated mean free path. It follows that
if the two types of interactions are statistically uncorrelated, then
the {\it effective mean free path} $\lambda$ is given by
$$
{1 \over \lambda} = {1 \over \lambdaii} + {1 \over \lambdacoh}\,,
\eqno(2.1)
$$
where $\lambdacoh$ is the mean distance between kinks in the field,
which is equivalent to the coherence or correlation length.

We will continue to focus on the case of a plane-parallel shear
flow characterized by the velocity distribution
$$
\vec u = u(y)\,\hat z\,, \eqno(2.2)
$$
where $u(0)=0$. To eliminate unnecessary complexity, we will also
assume that the ions are isothermal with temperature $T_i$. This
is reasonable so long as the temperature does not vary on scales
shorter than the ion effective mean free path $\lambda$. It will
be convenient to introduce a local polar coordinate system
$(r,\theta,\phi)$ using the standard transformation
$$
x = r\,\sin\theta \cos\phi\ \ \ \,,\ \ \ 
y = r\,\sin\theta \sin\phi\ \ \ \,,\ \ \ 
z = r\,\cos\theta\,, \eqno(2.3)
$$ 
in which case the velocity $v_r$ along the $\hat r$-direction is
related to $v_x$, $v_y$, and $v_z$ by
$$
v_r = v_x\,\sin\theta \cos\phi + v_y\,\sin\theta \sin\phi
+ v_z\,\cos\theta\,. \eqno(2.4)
$$

Let us first consider a case with no magnetic field. Then, viewed
from a frame comoving with the local fluid, the local ions have a
Maxwellian velocity distribution with temperature $T_i$. However,
viewed from the rest frame of the fluid located at $y=0$, the
distributions of $v_x$, $v_y$, and $v_z$ for particles located
at an arbitrary value of $y$ are given by
$$
f(v_x) = \left(m_i \over 2 \pi k T_i\right)^{1/2}
\exp\biggl\{-{m_i \over 2 k T_i} v_x^2 \biggr\}\,,
$$
$$
f(v_y) = \left(m_i \over 2 \pi k T_i\right)^{1/2}
\exp\biggl\{-{m_i \over 2 k T_i} v_y^2 \biggr\}\,,
$$
$$
f(v_z) = \left(m_i \over 2 \pi k T_i\right)^{1/2}
\exp\biggl\{-{m_i \over 2 k T_i} [v_z - u(y)]^2 \biggr\}\,,
\eqno(2.5)
$$
due to the presence of the shear flow, where $f(v_i)\,dv_i$ gives
the fraction of particles with $i^{\rm th}$ component of velocity
between $v_i$ and $v_i + dv_i$, and
$\int_{-\infty}^{\infty} f(v_i)\,dv_i=1$. Since $v_x$, $v_y$,
and $v_z$ are independent random variables, it follows from equation (2.4) that
the distribution of $v_r$ is given by
$$
f(v_r) = \left(m_i \over 2 \pi k T_i\right)^{1/2}
\exp\biggl\{-{m_i \over 2 k T_i}
[v_r - u(y) \cos\theta]^2 \biggr\}\,. \eqno(2.6)
$$
Next we consider the effect of ``turning on'' a magnetic field
oriented in the $\hat r$-direction specified by the angles
$(\theta,\phi)$. If the field is so strong that
$\lambdacyc \ll \lambdaii$, then the ions spiral tightly
around the field lines. However, the component of the velocity
parallel to the field ($v_r$) is completely unaffected, and therefore
the distribution of $v_r$ is still given by equation~(2.6) even
in the presence of a magnetic field.

We wish to compute the $\hat y$-directed flux of $\hat z$-momentum
due to particles crossing the $y = 0$ plane from both sides along the
 field line. It may be noted that since we assume $u(0) = 0$, layers on 
either side of this plane will have oppositely directed flow velocities. 
Since we expect that $\lambdacyc \ll \lambdaii$
in most cases of interest, we shall adopt the ``bead-on-string'' model
for the particle transport and work in the limit
$\lambdacyc / \lambdaii \to 0$, in which case $v_y$ and $v_z$ are
given by
$$
v_y = v_r\,\sin\theta \sin\phi\,,\ \ \ \ \ 
v_z = v_r\,\cos\theta\,. \eqno(2.7)
$$
Hence we ignore the components of momentum perpendicular to the
field and consider only the transport of momentum along the field lines.
For the purpose of calculating the momentum flux, it is sufficient
to consider particles starting out at a distance $\lambda$ from the
origin. It follows that at the starting point
$$
y = \lambda\,\sin\theta \sin\phi\,, \eqno(2.8)
$$
and therefore
$$
u(y) = u'(0)\,\lambda\,\sin\theta \sin\phi \eqno(2.9)
$$
to first order in $\lambda$, where the prime denotes differentiation
with respect to $y$. The $\hat y$-directed flux of $\hat z$-momentum
due to particles approaching the origin from both sides of the $y = 0$ 
plane is given by
$$
P(\theta,\phi) = 2\,\int_{-\infty}^0 [m_i v_z] \cdot
[N_i\,v_y\,f(v_r)\,dv_r]\,, \eqno(2.10)
$$
where the first term inside the integral is the $\hat z$-momentum
carried by the particles and the second term is the $\hat y$-directed
particle flux. Then to first order in $\lambda$ we obtain
$$
P(\theta,\phi) = 2\,m_i\,N_i\,\cos\theta \sin\theta \sin\phi
\left[{k T_i \over 2 m_i} - \left(2 k T_i \over\pi m_i\right)^{1/2}
u'(0)\,\lambda\,\cos\theta \sin\theta \sin\phi \right]\,, \eqno(2.11)
$$
which gives the shear stress as a function of $\theta$ and $\phi$.
The first term on the right-hand side describes the ``thermal stress''
due to the stochastic drifting of particles along the field lines, which
occurs even in the absence of a velocity gradient. The second term
gives the modification due to the presence of the velocity gradient.
Equation~(2.11) vanishes when the field is exactly perpendicular to
the velocity gradient ($\sin\theta \sin\phi=0$), which agrees with 
equation~(1.8) for the cross-field viscosity in the
limit $\lambdacyc / \lambdaii \to 0$.

Equation~(2.11) for the direction-dependent stress can be used to
construct two-dimensional models that treat both the radial and azimuthal
structure of the disk. In these models, the direction of the local magnetic
field is a random function of the radial and azimuthal position on scales
exceeding the coherence (correlation) length $\lambdacoh$. In order to
construct one-dimensional models, we need to average equation~(2.11) over
all directions to obtain the mean stress
$$
\langle P \rangle \equiv {1 \over 4 \pi} \int P(\theta,\phi)\,d\Omega\,,
\eqno(2.12)
$$
where $d\Omega = \sin\theta\,d\theta\,d\phi$, and
 $0 < \theta < \pi$, $0 < \phi < 2 \, \pi$. Substituting equation~(2.11) into
equation~(2.12) and integrating over $\theta$ and $\phi$ yields for the mean
(direction-averaged) hybrid viscosity
$$
\etahyb \equiv - {\langle P \rangle \over u'(0)}
= {2 \over 15}\,m_i\,N_i\,\lambda
\left(2 k T_i \over \pi m_i \right)^{1/2}\,. \eqno(2.13)
$$
Note that the ``thermal stress'' appearing in equation~(2.11) is
symmetric and therefore it vanishes upon integration.

We can also write the hybrid viscosity given by equation (2.13) as
$$
\etahyb = {2 \over 15}\,{\lambda \over \lambdaii}\,\etaff\,, \eqno(2.14)
$$
where $\etaff$ is the standard, field-free Coulomb viscosity given
by equation~(1.5). We see that no factor describing the efficiency of
the momentum transfer process appears in the expression for $\etahyb$,
in contrast to the cross-field viscosity $\eta_\perp$ given by
equation~(1.8). This is because in the hybrid case particles
originating on the right side of the plane and crossing over
definitely deposit their momentum on the left side. Since
$\etahyb / \etaff \sim (\lambda / \lambdaii)$ and
$\eta_\perp / \etaff \sim (\lambdacyc / \lambdaii)^2$, in
it is clear that the hybrid viscosity will
greatly exceed the cross-field viscosity if
$(\lambda / \lambdacyc) \gg (\lambdacyc / \lambdaii)$, which
is likely to be well satisfied in hot accretion disks, as will be seen 
in \S 3. This suggests that the ions play a much larger role in the
momentum transfer process in the presence of magnetic fields
than originally concluded by Paczynski (1978). In \S~3 we use
our results to analyze the structure of a two-temperature
quasi-Keplerian accretion disk with unsaturated inverse-Compton
cooling.

\bigskip
\centerline{\bf 3. APPLICATION TO TWO-TEMPERATURE ACCRETION DISKS}
\bigskip

We consider the two-temperature, steady-state model first
proposed by Shapiro, Lightman, \& and Eardley (1976) and
adopted by Eilek \& Kafatos (1983). In this model the ions
and electrons are coupled only via Coulomb collisions and
the electrons with temperature $T_e$ are assumed to radiate
their energy away via unsaturated inverse-Compton cooling.
In this case the two-temperature condition $T_i \gg T_e$
is satisfied if
$$
t_{\rm e-i} > t_{\rm accr} > t_{\rm ii} > t_{\rm ee} \,,
\eqno (3.1)
$$
where $t_{\rm ei}$, $t_{\rm ee}$, and $t_{\rm ii}$ are the timescales
for electron-ion, electron-electron, and ion-ion Coulomb equilibration,
respectively, and $t_{\rm accr}$ is the timescale for accretion onto
the black hole.
We will use the viscosity prescription given by equation (2.13),
and we will assume that the coupling between ions and electrons
occurs exclusively via Coulomb  
interactions. Hence we neglect the possibility
that collective plasma processes might result in an additional  
coupling between the ions and electrons, over and above the usual
Coulomb coupling, which could in principle lead to a violation of
the two-temperature condition. However, Begelman \& Chiueh (1988) considered
this possibility, and concluded that such collective processes are
not likely to strongly affect the thermal structure of the disk.
Equations~(A1--A6) in appendix A list the basic structure equations
for the two-temperature quasi-Keplerian disk model. Equations (A7)--(A9) in
appendix A constitute a list of the analytical solutions to these structure
equations, which are derived under the assumption that $T_i \gg T_e$.
These solutions have an arbitrary $\alpha$ parameter built into them,
which in general can be treated as a constant or allowed to vary with
radius using a specific model for the viscosity. In our case the
variation of $\alpha$ is obtained by substituting our expression for
$\etahyb$ into equation~(1.1).

In order to close the system of equations and obtain solutions for
the disk structure, we must also adopt a model for the variation of
the magnetic coherence length $\lambdacoh$ which appears in the
definition of the effective mean free path $\lambda$ (eq.~[2.1]).
We assume here that the field topology varies in a self-similar
manner with the local radius $R$, so that
$$
\lambdacoh \equiv \xi \, R\,, \eqno(3.2)
$$
where $\xi$ is a free parameter which we set equal to a constant for
a given model. It follows from the definition of $\lambda$ that
$$
{R \over \lambda} = {R\over\lambdaii} + {1\over \xi}\,,
\eqno(3.3)
$$
which implies that $\lambda/R \leq \xi$, with equality occurring in the
limit $\xi \to 0$. Imposing the restriction $\xi \leq 1$ (which is
inherent in the assumption of tangled magnetic fields) thus guarantees
that $\lambda/R \leq 1$, preserving the validity of the fluid
description of the plasma.

\bigskip
\centerline{3.1. \it A Two-Temperature Accretion Disk Model}
\bigskip

In a cylindrically symmetric accretion disk, the relevant component
 of the stress arising from the hybrid viscosity is given by
$$
\alpha_{\rm hyb} \, P \equiv - \etahyb R \, {d \Omega_{{\rm kepl}} 
\over dR} \, ,  
\eqno(3.4)
$$
which is equivalent to equation (1.1). We use equation (3.4)
to derive $\alpha_{\rm hyb}$ from $\etahyb$. Equations (A7)--(A9) in 
appendix A and equation (3.4) jointly yield  
the following self-consistent solutions for the model:
$$
\alpha = 147.31  \, \delta^{1/3} f_{1}^{-1/6} f_{2}^{2/3} \biggl 
({\dot{M}_{*} \over M_{8}}  
\biggr)^{2/3} \tau_{es}^{-1} R_{*}^{-1} \eqno (3.5)
$$
$$
T_{i} = 3.38 \times 10^{11} \, \delta^{-1/3} f_{1}^{1/6} f_{2}^{1/3}
 \biggl ({\dot{M}_{*} \over  
M_{8}} \biggr)^{1/3} R_{*}^{-1/2} \eqno (3.6)
$$
$$
T_{e} = {1.40 \times 10^{9} y \over \tau_{es} (1+\tau_{es})} \eqno (3.7)
$$
$$
N_i = 5.70 \times 10^{11} \, \delta^{1/6} f_{1}^{5/12} f_{2}^{-1/6} 
\biggl ({\dot{M}_{*} \over  
M_{8}} \biggr)^{5/6} \dot{M_{*}}^{-1} \tau_{es} \, R_{*}^{-5/4} \eqno (3.8)
$$
$$
{H \over R} = 0.175  \, \delta^{-1/6} f_{1}^{-5/12} f_{2}^{1/6}
 \biggl ({\dot{M}_{*} \over  
M_{8}} \biggr)^{1/6} R_{*}^{1/4} \,, \eqno (3.9)
$$
where
$$
\delta \equiv {\lambda \over \lambdaii}
= \left(1 + {\lambdaii \over \xi\,R} \right)^{-1} \,. \eqno(3.10)
$$
and
$$
\dot M_{*} \equiv {\dot{M} \over 1 \mrate}\, \, \, ,
M_8 \equiv {M \over 10^{8} \msun}\, \, \, , R_{*} \equiv {R \over GM/c^{2}}.
 \eqno(3.11)
$$

The following two equations jointly define an implicit algebraic equation 
for determining  
$\delta$ as a function of $R_{*}$ for given ($\xi$, $y$, 
${\dot{M_{*}} / M_{8}}$).
$$
\tau_{es} = 160.214 \, \xi^{-1} {\delta^{1/6} \over 1 - \delta} 
f_{1}^{-1/12} f_{2}^{5/6} \biggl  
({\dot{M}_{*} \over M_{8}} \biggr)^{5/6} R_{*}^{-3/4} \, \eqno (3.12)
$$
$$
\tau_{es}^{7/3} (1 + \tau_{es}) = 57.8819 \, \delta^{1/9} f_{1}^{-7/18}
 f_{2}^{-1/9} f_{3}^{2/3}  
y \biggl ({\dot{M}_{*} \over M_{8}} \biggr)^{5/9} R_{*}^{-5/6} \,
\eqno(3.13)
$$
 We will restrict our attention to $1 > \xi > 0$, since $\xi 
\gapprox 1$ implies that
the field is strongly ordered over macroscopic length scales, which violates
our assumption that the field is tangled.It can be
seen from the definitions of $\delta$ and $\lambda$ that $0 < \delta < 1$.
This simplifies the task of searching for a root for $\delta$. Once a
root for $\delta$ is determined for a given $\xi$, it is used in 
equation~(3.12),
and the result obtained for $\tau_{es}$ is then used in equations~(3.5--3.9) to
determine the disk structure. In principle, therefore, one could compute
a disk model for a given $y$ and any combination of $\xi$,  $\dot{M}_{*}$
 and $M_{8}$. We consider $0.001 < \dot{M}/\ME < 1$,
where $\ME=\LE/c^2$ and $\LE=4 \pi G M m_p c/\sig$ is the Eddington luminosity
and $\sig$ is the Thomson cross section. Note that 
$\ME = 0.22 \,\dot{M}_{*}/M_{8}$.
Accretion rates that are close to the Eddington value are more likely to
be significant from the point of view of observations.

\bigskip
\centerline{3.2. \it Model Self-Consistency Constraints}
\bigskip

For the models to be self-consistent, they have to fufill the following 
conditions:

(i) $\lambda/R < 1$. This assures us of the validity of applying the fluid
 approximation to  
the plasma. As discussed above, imposing $\xi < 1$ ensures the satisfaction of  
this criterion.

(ii) $T_{i}/T_{e} >> 1$. This is the essence of the two-temperature condition.
 Furthermore,  the analytical solutions given by equations (A7)--(A9) are 
valid only if this is true.

(iii) $H/R < 1$. This ensures that the disk remains geometrically thin. 
This is yet another  
condition that is assumed in deriving the analytical solutions listed in
 appendix A. As we shall see,  
this imposes the most severe restriction on achievable accretion rates.

(iv) Of the different kinds of viscosity that can possibly exist in the 
accretion disk, we assume the hybrid viscosity we have derived here to be the
 dominant form. The hydrodynamic turbulent viscosity used by 
Shakura \& Sunyaev (1973) is based on dimensional arguments, and, 
according to Schramkowski \& Torkelsson (1995), is probably less significant 
than viscosity arising from MHD turbulence, in which the magnetic field 
plays a significant role. We will discuss the contribution of what is
 referred to as pure magnetic viscosity (as opposed to our hybrid viscosity) 
in \S 4. For relatively
high accretion rates (close to the Eddington limit), one would expect rather 
high
luminosities. Consequently, the contribution of radiation viscosity, which 
is characterized by an associated $\alpha_{\rm rad}$,  would be
appreciable. Appendix B describes how
$\alpha_{\rm rad}$ is calculated. Since we are not including radiation 
viscosity in our treatment, we
need to remain in a region where $\alpha_{\rm hyb} > \alpha_{\rm rad}$, in order
to be self-consistent.

Figure 1 shows the nature of these restrictions for a maximally rotating 
Kerr black hole  
($a/M = 0.998$). Each point in the parameter space spanned by $\xi$ and 
$\Mdot/\ME$  
represents a potential model. It may be noted that each of these models 
assumes a  
constant value of $\xi \equiv {\lambda_{\rm coh}/ R}$ throughout the extent
 of the disk.  
This implies a certain self-similarity in the manner in which the embedded 
magnetic field is  
tangled. For each criterion (the $H/R < 1$ criterion, for instance), 
``allowed'' models are  
defined as those for which that criterion is satisfied throughout the
 disk ($R_{\rm ms} <  
R_{*} < 50$), where $R_{\rm ms}$ is the radius of marginal stability for the
 metric under consideration and $R_{*}$ is defined in equation (3.11). We 
take the entire region under consideration to be gas-pressure dominated 
(as in Shapiro, Lightman \& Eardley 1976), and arbitrarily take the outer 
boundary of the disk to be at $R_{*} = 50$. While we have verified that the 
gas pressure is indeed dominant over the radiation pressure in all cases of 
interest here, adopting an outer boundary of $R_{*} = 50$ is still an 
arbitrary measure. 

Since  
we have restricted ourselves to $\xi < 1$, condition (i) is automatically  
satisfied. It also turns out that the condition $T_{i}/T_{e} > 1$ is satsified throughout the  
parameter space shown in Figure 1. Constraints (iii) and (iv) are shown in Figure 1. Fully  
self-consistent models are possible only in the extreme left hand segment of the plot,
indicated by $H/R < 1$, $\alpha_{\rm hyb} > \alpha_{\rm rad}$. Evidently,
$H/R < 1$ is the  
condition that imposes the most severe restraint on achievable accretion rates. There is a  
small range of accretion rates and $\xi$, represented by the region between the two lines,  
where the disk might be puffy ($H/R > 1$), but the model is partially self-consistent in the  
sense that $\alpha_{\rm hyb} > \alpha_{\rm rad}$. On the extreme right hand side of the  
plot, the luminosity is high enough to cause $\alpha_{\rm rad}$ to be greater than $\alpha_{\rm hyb}$ and our models are no  
longer self-consistent.

Figure 2 illustrates the corresponding constraints for a Schwarzschild black hole. Schwarzschild black holes are in general cooler than Kerr black holes, because the radius of marginal stability is greater. Therefore the accretion disks around these objects are apt to be less puffy. This is reflected in the  
relatively more benign $H/R$ constraint in Figure 2. It also turns out that, owing to relatively lower luminosities, the $\alpha_{\rm hyb} > \alpha_{\rm rad}$ constraint is somewhat more forgiving for the Schwarzschild case, allowing relatively higher accretion rates to be achieved. For both the Kerr and Schwarzschild cases, it is seen from Figures 1 and 2 that the self-consistency requirements for our model impose an upper bound on the maximum attainable accretion rate $\Mdot/\ME$. However, high accretion rates are of interest from the point of view of observations, since they result in high luminosities and are therefore relatively easier to detect. Hence, models with the highest possible accretion rates allowed by the self-consistency constraints discussed above are likely to be significant from the point of view of observations.

We now examine a typical model from each of the parameter spaces illustrated in Figures 1  
and 2. We choose an accretion rate of $\Mdot/\ME = 0.5$ and a constant value of $\xi =  0.8$ throughout the flow for both the Kerr and Schwarzschild cases. It can be seen from  
Figures 1 and 2 that these values will ensure that both the Kerr and Schwarzschild models  
will fulfill all the self-consistency requirements. Figures 3--5 represent the Kerr model with  
$\Mdot/\ME = 0.5$ and $\xi = 0.8$, while Figures 6--8 represent the Schwarzschild model  
with the same parameters. The relatively high ion temperatures in both cases are worth  
noting, and is indicative of the fact that these models are good candidates for the  
production of high energy gamma rays (Eilek \& Kafatos 1983, Eilek 1980). Furthermore, as noted earlier, the Schwarzschild  
disk is relatively cooler and consequently less puffy than the Kerr disk. Except for the  
region near the radius of marginal stability, $\alpha_{\rm hyb}$ is seen to be nearly  
constant in the Kerr case. This indicates that as far as this particular viscosity mechanism  
is concerned, the assumption of a constant $\alpha$ taking on values ranging from 0.01 to  
0.1 (as adopted by the standard disk model of Eilek \& Kafatos 1983) is quite good.  
Closer scrutiny of the region near the radius of marginal stability reveals that the $\alpha_{\rm hyb}$  
and $\tau_{\rm es}$ curves in the Kerr case are in fact continuous. The rapid increase in  
$\alpha_{\rm hyb}$ is due primarily to the drop in pressure caused by the decrease in  
temperature near that region, while the sharp drop can be attributed to behavior of the  
relativistic correction factors $f_{1}$, $f_{2}$ and $f_{3}$ near the radius of marginal  
stability.

\bigskip
\centerline{\bf 4. DISCUSSION}
\bigskip

We have derived a hybrid viscosity arising from momentum deposition by ions in the  
presence of a tangled magnetic field. This viscosity is neither the usual Coulomb viscosity  
which arises from Coulomb collisions between ions, nor is it pure magnetic viscosity, which is due  
to magnetic stresses. The tangled magnetic field plays a role in confining the ions, which makes the viscosity mechanism a local process. The field also acts as an intermediary in the momentum transfer between ions, in situations where the coherence length of the field $\lambda_{\rm coh} \ll \lambda_{\rm ii}$, where $\lambda_{\rm ii}$ is the usual Coulomb ion-ion mean free path. Upon application of this form of viscosity to a specific disk model, we observe that the self-consistency requirements limit valid models to sub-Eddington accretion rates. Otherwise, the disks become puffy and radiation viscosity dominates over hybrid viscosity. This could be interpreted as a statement favoring the possibility of quasi-spherical, radiation viscosity-supported accretion for near-Eddington accretion rates.

The Shakura Sunyaev $\alpha$ parameter arising from this hybrid viscosity, $\alpha_{\rm hyb}$, is seen to lie  
roughly between 0.01 and 0.1. It is interesting to compare this with the values of $\alpha$  
one would expect to obtain from pure magnetic viscosity i.e; that arising from magnetic  
stresses alone. There have been several attempts at quantifying magnetic viscosity by  
detailed computations of the magnetic field arising from dynamo processes (Eardley \&  
Lightman 1975, for instance). If we consider the magnetic stress to be equal to the magnetic pressure $P_{B} = B^{2}/(8 \pi)$, the $\alpha$ parameter arising out of pure magnetic viscosity is defined by
$$
\alpha_{\rm mag} \, P \equiv {B^{2} \over 8 \pi} \,. \eqno(4.1)
$$
where $P$ is the total pressure. If we define $\beta = P_{g}/P_{B}$, where $P_{g} = N_{i} k (T_{i} + T_{e})$, $\alpha_{{\rm mag}} \sim 1/(\beta + 1)$. The  
value of $\alpha_{\rm mag}$ can thus vary from 0.5 at $\beta =  
1$ (equipartition) to as low as $\sim 0.01$ for $\beta = 100$. The values of $\alpha_{\rm hyb}$ we obtain are thus seen to be  
comparable to those obtained from pure magnetic viscosity. However, the magnitude of the hybrid viscosity is quite insensitive to the magnitude of the magnetic field, unlike the  
situation with pure magnetic viscosity. The only restriction on the magnitude of the magnetic  
field in our calculations is that it be at least so large as to warrant the assumption of nearly  
zero gyroradii for the ions.
Our calculations neglect any finite gyroradius effects, and in effect consider a lower limit  
on the possible momentum transfer. It is, however, a realistic one, since a magnetic field  
as weak as $\sim 10^{-7}$ Gauss (which corresponds to a rather large $\beta$, very far  
below equipartition) is sufficient to cause the gyroradius to be smaller than any of the  
macroscopic disk dimensions. It is quite likely that magnetic fields much larger than that  
value, and much closer to the equipartition value, will be embedded in the accreting plasma.

We have entirely neglected any momentum transfer arising out of short wavelength  
plasma waves in the accretion flow. Since the tangled magnetic field is taken to be arising  
from plasma turbulence, the presence of such waves is quite plausible, and it is one  
aspect of the problem we have neglected in our calculations. One could model an  
ensemble of such turbulent plasma waves as a collection of plasmons, assign a number  
density and mass to these entities and investigate their role as intermediaries in  
momentum transfer. A self-consistent calculation of the tangled magnetic fields arising as  
a consequence of the presence of plasma turbulence could also reveal magnetic flutter;  
temporal variations in the local magnetic field (as distinct from the large scale evolution of  
the fields due to dynamo action that we have discussed) that we have also neglected.

We now turn our attention to the deficiencies in our treatment of the disk structure. Our  
calculations are time-independent; they assume the presence of a steady state. This might  
or might not be true, and there have been a number of investigations of possible disk  
instabilities (Shakura \& Sunyaev 1976, Piran 1978, for instance) which consider the presence of thermal and viscous instabilities that could   
break up the disk and cause variations in the disk luminosity. The temperature-dependent  
nature of any viscosity in which ions play a part (like the hybrid viscosity discussed in this  
paper) would result in a coupling of viscous and thermal instabilities. The presence of  
magnetic fields can be expected to stabilize possible instabilities arising out of the cooling  
mechanisms, but it is not clear if it will have any effect upon instabilities in the viscous  
heating rate. We are currently in the process of undertaking an investigation of these  
aspects.

We emphasize that we do not make any attempt to self-consistently calculate the  
topology of the tangled magnetic field. Instead, we merely use $\xi \equiv \lambda_{\rm coh}/R$ as a parameter. The  
presence of the tangled magnetic field serves the following purposes:

(i) The ions are effectively caught in the ``net'' of tangled magnetic field lines, and (since  
$\xi < 1$) are confined to remain well within the accretion flow. The tangled magnetic field  
cannot alter the net energy of the distribution, and we therefore assume the temperature of  
the distribution to be the same as what it would have been in the absence of the magnetic  
field. In effect, we assume that the ions can still relax to a Maxwellian distribution at the  
same temperature, although we have not rigorously investigated the relaxation time  
associated with such a situation.

(ii) Our analysis assumes that the ions traveling from one layer to another along a field line  
see no temporal variations in the magnetic field. If we assume the magnetic field to evolve  
(due to dynamo action) over timescales comparable to the Keplerian timescale (orbital period), we can  
write
$$
{\Delta t \over t_{B}} \simeq [{\lambda_{\rm coh} \over v_{i}}]/[{R \over v_{{\rm kepl}}}] = \xi {v_{i}  
\over v_{{\rm kepl}}} \,. \eqno (4.2)
$$
where $\Delta t$ denotes the time for an ion to travel from one layer to another that is  
separated by a distance $\simeq \lambda_{\rm coh}$, $t_{B}$ denotes the timescale over  
which the magnetic field evolves and $v_{i}$ is the ion thermal speed.
Since we are restricted to $\xi < 1$ and we expect $v_{i}/v_{{\rm kepl}} \leq 1$, it is  
evident from equation (4.2) that the aforementioned assumption is quite valid.

To emphasize the main points, we have derived a new kind of viscosity called ``hybrid'' viscosity and have shown that ions play a much more important role than previously thought in transporting angular momentum in accretion disks that have tangled magnetic fields embedded in them. Although we have not modeled the way in which these magnetic fields are dynamically generated, it is quite plausible that small scale MHD turbulence in the accretion flow can give rise to such fields. We have also considered this form of viscosity vis-a-vis other forms of viscosity that can exist in such situations. The temperature-dependent nature of this ``hybrid'' viscosity also makes the study of possible instabilities in the disk a very interesting and relevant question to be resolved.

\bigskip
\centerline{APPENDIX A}
\centerline{CONSTITUTIVE EQUATIONS FOR TWO-TEMPERATURE, COMPTONIZED MODEL}
\bigskip

The basic disk structure equations are the same as those used in the disk structure  
calculations of Eilek \& Kafatos (1983), which neglect radiation pressure:
$$
P = {G M m_{i} N_i H^{2}f_{1} \over R^{3}} \, ,\eqno(\rm A1)
$$
$$
{\alpha} P = {(G M R)^{{1 \over 2}} \dot{M} f_{2} \over 4 \pi R^{2} H} \, ,
\eqno(\rm A2)
$$
$$
{3 \over 8 \pi} {G M \dot{M} \over R^{3} H} f_{3} = 3.75 \times 10^{21} \,
m_{i} \ln \Lambda N_i^{2} k {(T_{i} - T_{e}) \over T_{e}^{{3 \over 2}}} \, ,
\eqno(\rm A3)
$$
$$
P = N_i k (T_{i} + T_{e}) \, ,\eqno(\rm A4)
$$
$$
T_{e} = {m_{e} c^{2} y \over 4 k} {1 \over \tau_{es} g(\tau_{es})} \, ,
\eqno(\rm A5)
$$
$$
\tau_{es} = N_i \sig H \, ,\eqno(\rm A6)
$$
where $\sig$ is the Thomson scattering cross-section.
The Coulomb logarithm $\ln \Lambda$ is taken to be 15 in our numerical calculations and the function $g(\tau_{es}) \equiv 1 + \tau_{es}$. It may be noted  
that this is different from the form for $g(\tau_{es})$ used by Eilek \& Kafatos (1983). 
$f_{1}$, $f_{2}$ and $f_{3}$ are the relativistic correction factors appropriate to the metric  
under consideration. These factors for a Kerr black hole with ${a / M} = 0.998$ are used in  
Eilek \& Kafatos (1983). Eilek (1980) gives plots of $f_{1}$, $f_{2}$ and $f_{3}$ for a Kerr  
black hole. The relativistic correction factors appropriate to a  
Schwarzschild metric can be obtained by setting $a/M = 0$ in the expressions for $f_{1}$,  
$f_{2}$ and $f_{3}$. In keeping with the convention used in Eilek \& Kafatos (1983), we make the definitions 
$M_{8} \equiv M/10^{8} \msun$, $\dot{M}_{*} \equiv \dot{M}/1 \mrate$, and $R_{*} \equiv R/(GM/c^{2})$.  
If we assume $T_{i} >> T_{e}$, equations (A1)--(A6) yield the following analytical solutions:
$$
T_{i} = 4.99 \times 10^{13} {\dot{M}_{*} \over M_{8}} f_{2} \tau_{es}^{-1} \alpha^{-1}  
R_{*}^{-3/2} \, ,\eqno(\rm A7)
$$
$$
T_{e} = 1.40 \times 10^{9} y \, \tau_{es}^{-1} \bigl ( g(\tau_{es}) \bigr
)^{-1} \, ,\eqno(\rm A8)
$$
$$
N_i = 4.70 \times 10^{10} \biggl ( {\dot{M}_{*} \over M_{8}} \biggr )^{1/2}
\dot{M}_{*}^{-1}f_{1}^{1/2} f_{2}^{-1/2} \tau_{es}^{3/2} \alpha^{1/2} R_{*}^{-3/4}\,.
\eqno(\rm A9)
$$

It may be emphasized that $\alpha$ is a free parameter in the above solutions.

\bigskip
\centerline{APPENDIX B}
\centerline{DEFINITION OF RADIATION VISCOSITY}
\bigskip

We use an $\alpha_{\rm rad}$, the $\alpha$ parameter obtained from radiation viscosity, as  
a diagnostic in this paper. We now proceed to define the manner in which we compute  
$\alpha_{\rm rad}$. We follow Shapiro, Lightman \& Eardley (1976) in defining the  
radiation energy density using
$$
U_{\rm rad} = (F/c) \, g(\tau_{es}) \, . \eqno(B1)
$$
 If the $y$ parameter is taken to be equal to unity, eliminating $g(\tau_{es})$ between equations (A5) and (B1), using equation (A6) yields
$$
{F \over H} = \biggl ( {4 k T_{e} \over m_{e} c^{2}} \biggr ) N_i \sig c U_{\rm rad}  
\, . \eqno(\rm B2)
$$
where $F$ is the dissipated energy density. Equation (A3) is another way of defining ${F /  
H}$; in fact, $F$ has to be equal to $\bigl (3/8 \pi \bigr ) G M \Mdot/R^{3}$ for the disk to be  
quasi-Keplerian. Equating the right hand side of equation (A3) to that of equation (A8) and  
assuming $T_{i} >> T_{e}$ yields
$$
U_{\rm rad} = 9.565 \times 10^{5} N_i  \, T_{i} \, T_{e}^{-5/2} \,
. \eqno(\rm B3)
$$
We next adopt the definition of radiation viscosity $\eta_{\rm rad}$ given by Loeb and Laor  
(1992),
$$
\eta_{\rm rad} = {8 \over 27} {U_{\rm rad} \over N_i \sig c} \,
. \eqno(\rm B4)
$$
We calculate $\alpha_{\rm rad}$ by adopting the usual definition for $\alpha$, akin to  
equation (3.4),
$$
\alpha_{\rm rad} \, P \equiv - R \, \eta_{\rm rad} {d \Omega_{\rm kepl} \over dR} \,
. \eqno(\rm B5)
$$
It may be noted that we assume the disk to be gas-pressure dominated;  
although we do calculate $\alpha_{\rm rad}$ as a diagnostic tool, $P$ in equation (B4), which represents total pressure, does not include radiation pressure in our calculations.

This yields

$$
{\alpha_{\rm rad} \over \alpha_{\rm hyb}} = {\eta_{\rm rad} \over \eta_{\rm hyb}} = 6.45 \times 10^{33} \, \delta^{-1} \, \ln \Lambda \, T_{i}^{-3/2} T_{e}^{-5/2} \, , \eqno(\rm B6)
$$
where $\delta$ is defined in equation (3.10).

\centerline{REFERENCES}

\refs{

Begelman, M. C., \& Chiueh, T. 1988, ApJ,  332, 872

Eardley, D. M., \& Lightman, A. P. 1975, ApJ, 200, 187

Eilek, J. A. 1980, ApJ, 236, 644

Eilek, J. A., \& Kafatos, M. 1983, ApJ, 271, 804

Filho, C. M. 1995, A \& A, 294, 295

Goldman, I., \& Wandel, A. 1995, ApJ, 443, 187

Kafatos, M. 1988, in  Supermassive Black Holes (Cambridge University Press), 307

Kaufman, A., 1960, La Theorie dez Gas Neutres et Ionizes (Paris: Hermann et Cie)

Loeb, A., \& Laor, A. 1992, ApJ, 384, 115

Matsumoto, R., \& Tajima, T. 1995, ApJ, 445, 767

Mihalas, D., \& Mihalas, B. W. 1984, Foundations of Radiation Hydrodynamics (New York: Oxford University Press)

Paczynski, B. 1978, Acta Astronomica, 28, 253

Piran, T. 1978, ApJ, 221, 652

Pringle, J. E. 1981, ARA\&A, 19, 137

Schramkowski, G. P., \& Torkelsson, U. 1996, A\&AR, in press

Shapiro, S. L., Lightman, A. L., \& Eardley, D. M. 1976, ApJ, 204, 187

Shakura, N. I., \& Sunyaev, R. A. 1973, A\&A, 24, 337

Shakura, N. I., \& Sunyaev, R. A. 1976, MNRAS, 175, 613

Spitzer, L. 1962, Physics of Fully Ionized Gases (Interscience, New York)

}

\centerline{FIGURE CAPTIONS: Please e-mail psubrama@gmu.edu for figures}
FIG 1.---The ($\xi$, $\Mdot/\ME$) parameter space for the canonical Kerr metric with $a/M = 0.998$. Each point in the parameter space represents a potential model. The lines demarcate regions in which different constraints are fulfilled. The extreme left section of the plot in which $H/R < 1$ and $\alpha_{\rm rad}/\alpha_{\rm hyb} < 1$ is the one in which the models are fully self-consistent.

FIG 2.---The analog of Fig. 1 for the Schwarzschild metric.

FIG 3.--- Curves of $\alpha_{\rm hyb}$ and $\tau_{es}$ for a specific model in the Kerr metric, with $\Mdot/\ME = 0.5$ and $\xi \equiv \lambda_{\rm coh}/R = 0.8$. A more detailed inspection of the curves in Fig. 3 reveals they are in fact continuous. The reasons for the steep gradients near the radius of marginal stability are explained in the text.

FIG 4.---Log($T_{i}$) and Log($T_{e}$) for the model shown in Fig. 3. ($\Mdot/\ME = 0.5$, $\xi = 0.8$)

FIG 5.---$H/R$ and $\lambda/R$ for the model shown in Figs. 3 and 4. ($\Mdot/\ME = 0.5$, $\xi = 0.8$)

FIG 6.---Curves of $\alpha_{\rm hyb}$ and $\tau_{es}$ in the Schwarzschild metric, with ($\Mdot/\ME = 0.5$, $\xi = 0.8$)

FIG 7.---Log($T_{i}$) and Log($T_{e}$) for the model shown in Fig. 6. (Schwarzschild metric; $\Mdot/\ME = 0.5$, $\xi = 0.8$)

FIG 8.---$H/R$ and $\lambda/R$ for the model shown in Figs. 6 and 7. (Schwarzschild metric; $\Mdot/\ME = 0.5$, $\xi = 0.8$)

\end